\pdfoutput=1
\documentclass[doublecol]{epl2} 
\usepackage{amsmath,amssymb,graphicx}
\newcommand{\nn}{\nonumber }

\usepackage{color}

\definecolor{db}{cmyk}{1,0.4,0,0.4}
\definecolor{dr}{cmyk}{0,0.9,0.7,0.4}
\definecolor{dg}{cmyk}{1,0,1,0.2}

    \let\e=\varepsilon

\def\nn{\nonumber}

\def\be{\begin{equation}}
\def\ee{\end{equation}}
\def\bea{\begin{eqnarray}}
\def\eea{\end{eqnarray}}
\def\ba{\begin{array}}
\def\ea{\end{array}}

\begin{document}

\title{Magnetoresistance and localization in bosonic insulators}
\author{Markus M\"uller}
\institute{The Abdus Salam International Center for Theoretical Physics, Strada Costiera 11, 34151 Trieste, Italy.}

%\date{\today\\}

%71.55.Jv}

\abstract{
We study the strong localization of hard core bosons. Using a locator expansion we find that in the insulator, unlike for typical fermion problems, nearly all low-energy scattering paths come with positive amplitudes and hence interfere constructively. As a consequence, the localization length of bosonic excitations shrinks when the constructive interference is suppressed by a magnetic field, entailing an exponentially large positive magnetoresistance, opposite to and significantly stronger than the analogous effect in fermions. Within the forward scattering approximation, we find that the lowest energy excitations are the most delocalized. A similar analysis applied to random field Ising models suggests that the ordering transition is due to a delocalization initiated at zero energy rather than due to the closure of a mobility gap in the paramagnet. 
%The superfluid transition can be approached on a Bethe lattice with large connectivity, where the superfluid is found to emerge at the chemical potential, . 
}

\maketitle

Most disordered insulators display some form of variable range hopping transport~\cite{ESbook}, reflecting the localization of carriers at low energies~\cite{Anderson}. Naively, one might expect that in such insulators the quantum statistics of carriers is irrelevant, as particles essentially never exchange their places. 
However, when the hopping length becomes much larger than the distance between impurities, transport is in fact very sensitive to statistics, as probed, e.g. by orbital magnetic fields. For {\em fermions} the latter suppresses the destructive interference among alternative virtual paths, leading to a strong negative magnetoresistance~\cite{NSS, SivanImry, Zhao, SpivakShklovskiiReview,MedinaKardar,KardarBook}. This is a very non-trivial %remarkable 
manifestation of quantum interference in impurity bands. In this Letter we address the {\em bosonic} counterpart of this effect, which, remarkably, has always the opposite sign. The quantum statistics also manifests itself in a non-trivial energy dependence of localization and in the way delocalization is approached.    
%We also investigate how bosons approach the transition towards a delocalized superfluid, and highlight various ways in which  affects localization properties of the insulator.

The present study of disordered bosons is motivated by a variety of experimental situations involving bosonic insulators, such as in Josephson junction arrays, certain superconducting films, turned insulating by strong disorder, repulsive cold bosonic atoms in speckle potentials and artificial gauge fields~\cite{Roati,Dalibard}, helium in porous media or random quantum magnets.~\cite{IoffeMezard}. In the presence of strong disorder, the respective insulators are expected to be Bose glasses~\cite{GiamarchiSchulz}, whose low energy excitations are localized by disorder, but do not exhibit a spectral gap. Transport of such strongly disordered bosons is still scarcely studied, but poses a variety of interesting conceptual questions, which are not fully resolved yet~\cite{Mueller09, Aleiner09, IoffeMezard}. %The peak in magnetoresistance is one of the aspects awaiting a full theoretical explanation.

A particularly interesting aspect of localization is the magnetoresistance in charged {\em bosonic} insulators. Recent experiments in strongly disordered, superconducting InO$_x$ films~\cite{GantmakherInOxpeak, Shahar} have shown that a magnetic field not only destroys rapidly the already weak superconductivity~\cite{Hebard}, but also induces a giant positive magnetoresistance in the ensuing insulating state. Similar effects in magnetoresistance have been reported in amorphous films of TiN~\cite{TiN_MR}, Bi~\cite{Goldman2011}, and in patterned films~\cite{Valles},  c.f., the review~\cite{Gantmakher}. The giant positive magnetoresistance in the vicinity of the superconducting transition is intriguing. Mechanisms such as shrinking impurity wavefunctions or spin blocking of weakly interacting electrons, which may play a role in semiconductors~\cite{ESbook}, hardly apply to these systems~\cite{Mueller09}. Instead, experimental observations in transport~\cite{Hebard, GantmakherInOxpeak, Shahar, TiN_MR, Valles, Kapitulnik, TanParendo,LittleParks} and tunneling microscopy~\cite{SacepeNatPhys, Sacepe}, as well as theoretical model studies~\cite{Nandini, FeigelmanIoffeKravtsov, Nattermann} suggest the  importance of remnant electron pairing in the insulator, despite the absence of global phase coherence~\cite{MPAFisher90}. 
%In the presence of strong disorder, the resulting insulator is expected to be a Bose glass~\cite{GiamarchiSchulz}, whose low energy excitations are localized by disorder, but do not exhibit a spectral gap. Transport in such systems poses a variety of interesting conceptual questions, which are not fully resolved yet~\cite{Mueller09, Aleiner09, IoffeMezard}. %The peak in magnetoresistance is one of the aspects awaiting a full theoretical explanation.

While it is natural to expect a magnetic field to increase the resistance of a bosonic insulator, in continuation of its destructive effect on superconductivity, there is no satisfactory microscopic explanation of the giant effects seen in experiments yet, despite attempts at phenomenological explanations~\cite{Meir} or model calculations for granular systems~\cite{Beloborodov,Fistul}. The latter do not account for the fact that in the experimental films~\cite{Shahar} no well-defined granular structure exists, and the spectral gap for pairs is expected to be washed out by strong disorder.~\cite{Nandini} This suggests that, most likely, it is Cooper pair (boson) localization due to disorder, which induces the insulating behavior, rather than the opening of a homogeneous gap in the insulator.~\cite{Mueller09}

Here we study a microscopic model of an insulator of hard core bosons, subject to strong disorder potentials. This captures, e.g., electronic systems with a strong local electron pairing. By contrasting this model with similar fermionic models, we reveal the specific role of quantum statistics. As the simplest model containing all relevant ingredients we consider a lattice, whose sites can accommodate only one quantum particle due to strong onsite repulsion. For spinless fermions this is simply the non-interacting Anderson model for single particle localization~\cite{Anderson,NSS}. For hard core bosons the model was introduced by Ma and Lee \cite{MaLee} who considered disordered superconductors in terms of preformed pairs (Anderson pseudospins). This is a faithful low energy representation of single-band Hubbard models with a strong negative $U$ attraction.~\cite{CarlsonGubernatis}
%The model was recently revisited in \cite{IoffeMezard, Mueller09}.
Similar models were recently studied in Refs.~\cite{IoffeMezard, Syzranov}, using approaches based on large lattice connectivity. 
Our calculation scheme below can easily be generalized to grains or islands hosting many particles, as long as the charge gap on typical grains is much bigger than the hopping amplitude between grains. 

%attributed the magnetoresistance peak to the survival of  superconducting islands that have field-dependent properties, postulating a Coulomb blockade that affects only electrons, but not pairs. Ref.~\cite{Beloborodov} considered an array of normal and superconducting grains, where the field tunes the gaps of pairs and electrons in opposite directions, yielding an magnetoresistance peak.
%Ref.~\cite{Fistul} instead interpreted the peak as reflecting the first oscillation of the charging energy of a clean array of  Josephson islands in a magnetic field.
%However, the models~\cite{Beloborodov, Fistul} hardly do justice to the complexity of the experimental systems~\cite{Shahar} in which no well-defined granular structure exists, and the spectral gap for pairs should be washed out by strong disorder.~\cite{Nandini} The latter suggests that it is rather pair localization due to disorder, which induces the insulating behavior.~\cite{Mueller09}   

% based on grains and containing disorder only in the nature of the grains, may apply to the experiments
 
%\cite{AndersonPseudospins} pseudospins
  
\begin{figure}[h]
\centerline{\includegraphics[width=0.48\textwidth]{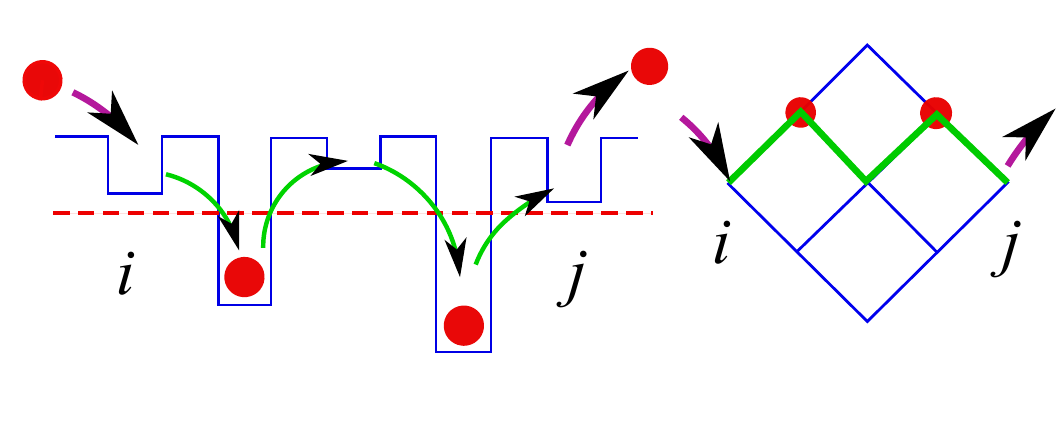}}
\caption{
In a coherent hopping process many particles move by one slot to the next negative energy site (the process shown in the left pannel corresponds to the highlighted paths on the right pannel). The many-body nature of this process is responsible for the statistical sign difference between bosons and fermions.
To leading order in the hopping, many alternative paths interfere in the Green's function $G^R_{i0}$ between sites $0, i$ (6 paths in the right pannel).
The sign of fermion amplitudes depends on the number of occupied sites (indicated by filled circles) on each path whereas bosonic paths all have positive amplitudes at low energy. A magnetic field suppresses the maximally constructive interference of bosons.}
\label{Fig:Interference}
\end{figure}

We consider a lattice of sites $i$ with random energies $\e_i$, uniformly distributed in $[-W,W]$, and weakly coupled by a  tunneling amplitude $t_{ij}= t$ between nearest neighbors,
\bea
\label{H}
H =%&=&H_0+H_1\\
%H_0 &=& 
\sum_i \e_i n_i %\\
%H_1 &=& 
- \sum_{\langle i,j\rangle }  t_{ij} (b^\dagger_j b_i +b^\dagger_i b_j),\quad n_i = b^\dagger_i b_i. 
\eea
$b_i^\dagger, b_i$ are creation and annihilation operators of fermions or hard core bosons, respectively. They    satisfy $b_i^2=0$, and the commutation relations $[b_i,b_j]_B=0$, $[b_i^\dagger,b_j]_B=\delta_{ij} (-1)^{B(1-n_i)}$, where 
$[.,.]_{B}$ denotes the commutator for bosons ($B=1$) and the anticommutator for fermions ($B=0$), respectively.~\cite{footnote:spinhalf}
In the presence of a magnetic field, the hopping acquires a phase $t_{ij} =t e^{-i \phi_{ij}}$, 
%**
the sum of $\phi_{ij}$ around a plaquette being proportional to the flux threading it. 
%where $q$ is the charge of the particle and $\phi_{ij} =\int_{r_i}^{r_j} \vec{A}\cdot d\vec{r}$ is evaluated along a straight path between site $i$ and $j$, and we set $\hbar =1$.

The important role of quantum statistics on magnetoresistance was noted early on by Zhao et al.~\cite{Zhao}, where low energy excitations were discussed. Here, we introduce an efficient new formalism, which allows us to give a rigorous derivation of their prediction, generalize it to finite energy excitations and gain insight on bosonic delocalization. The formalism is easily extendible to treat subleading corrections~\cite{Bapst}, and can be applied to many other disordered systems as well, as we will exemplify on the Ising model in random transverse fields.

We focus on the strongly insulating regime $t \ll W$. 
In the limit $t=0$ elementary excitations correspond to the addition or removal of a particle on given sites. For small hopping $t/W\ll 1$, these adiabatically deform into dressed excitations, which are still well localized in space. %In fact, one expects that {\em all} low energy excitations remain discrete and localized in this limit~\cite{Mueller09, IoffeMezard, Aleiner09}. % up to a certain threshold in energy, which may even scale with the volume.
The spatial properties of such many body excitations are well captured by the retarded Green's function
\bea
G^R_{i,0}(t-t') = -i\Theta(t-t') \langle [b_i(t),b_0^\dagger(t')]_B \rangle,
\eea
where $A(t)= e^{iHt} A(0)e^{-iHt}$.
It describes the amplitude of finding an extra particle at site $i$, after a time $t$ of adding a particle on site $0$.  To characterize the spatial decay of an excitation of given energy, one preferably works in frequency space,
$
G^R_{i,0}(\omega)= \int_{-\infty}^\infty G^R_{i,0}(t)e^{i\omega t} dt,
$
and extracting the relevant pole. Having hopping conductivity in mind, we are interested in low energy excitations, $\omega \ll W$.

We now analyze $G^R_{i,0}$ perturbatively in $t_{ij}$, which is justified deep in the insulator. 
Similarly as in early works of the Hubbard model~\cite{
%Zubarev, 
Hubbard}, we study the equation of motion 
\bea
\label{Eqmot}
i\frac{d}{dt} G^R_{i,0}(t-t') &=& \delta(t-t') \delta_{i,0}  \langle [b_0(0),b_0^\dagger(0)]_B \rangle \\
%&& -i \Theta(t-t') \langle [[b_i(t),H],b_0^\dagger(t')]_B \rangle.\nn\\
&& -i \Theta(t-t') \langle [i\dot{b}_i(t),b_0^\dagger(t')]_B\rangle ,\nn
\eea   
as a convenient starting point for a locator expansion in powers of the hopping $t/W$~\cite{Anderson}. This 
technique can easily be generalized to analyze other random field systems, too. It is easy to show that %for fermions
%\bea
%[b_i(t),H] =\e_i b_i(t) +\sum_{j \in \partial i} t_{ij} b_j(t),
%\eea
%while for hard core bosons 
\bea
\label{bdot}
%[b_i(t),H] = \e_i b_i(t) - (-1)^{B n_i(t)} \sum_{j \in \partial i} t_{ij}  b_j(t),\\
i\dot{b}_i(t)=[b_i(t),H] = \e_i b_i(t) - (-1)^{B n_i(t)} \sum_{j \in \partial i} t_{ij}  b_j(t),
\eea
where the sum runs over all neighboring sites of $i$. 
%We are interested in the decay of the correlation function at large distance. 

To leading order in $t/W$, we can restrict ourselves to forward scattering paths, in analogy to the fermionic (single particle) study by Nguyen et al.~\cite{NSS}. Hence, in Eq.~(\ref{bdot}) we retain only the neighbors $j$, which are closest to $0$, cf.~Fig.~\ref{Fig:Interference}. To leading order in $t/W$ we can decouple the sign factor in (\ref{bdot}) and use $\langle (-1)^{n_i(t)}...\rangle = {\rm sign}(\e_i)\langle ...\rangle +O((t/W)^2)$ in Eq.~(\ref{Eqmot})
%The frequency dependence will be discussed further below.
%As we will see below, for bosons this decay rate has a strong frequency dependence, unlike fermions .  %It is worthwhile stressing that this localization length is in general significantly larger  than the size of bosonic tight-binding  wavefunctions sitting on the sites $i$. In particular, 
to obtain the recursion relation
\bea
\label{recursion}
G^R_{i,0}(\omega) \approx \sum_{j\in \partial i,{\rm dist}(j,0) <{\rm dist}(i,0)} \frac{t_{ij}\, [{\rm sign}(\e_i)]^B}{\e_i-\omega} G^R_{j,0}(\omega).
\eea
This is easy to evaluate by a transfer matrix computation.  
Upon iteration of this forward scattering approximation, we obtain $G^R_{i,0}$ as a sum over all shortest paths ${\cal P}$ (of length $\ell$) between sites $0$ and $i$, which is {\em exact} to leading order in $t/W$,%(with $G^R_{00}(\omega)= [{\rm sgn}(\e_0)]^{B}/(\omega-\e_0)$),
%\bea
%(E-\e_i) G^R_{i,0}(E) = \sum_{j\in \partial i,{\rm dist}(j,0) ={\rm dist}(i,0)-1} V_{ij} (-{\rm sign}(\e_i))^{(1-F)} G^R_{j,0}(E)
%\eea
%Upon iteration, and using $G^R_{0,0}(E)=1/(E-\e_0)+O(V^2)$ (check sign), we find the final result
\bea
\label{result}
\frac{G^R_{i,0}(\omega)}{G^R_{0,0}(\omega)} = \sum_{{\cal P}=\{j_0=0,...,j_\ell=i \}} \prod_{p=1}^{\ell } \frac{t_{j_{p-1},j_p} [{\rm sgn}(\e_{j_p})]^{B} }{\e_{j_p}-\omega} .
\eea
By setting $\omega\to  \e_0$ and extracting the residue of the corresponding pole in $G^R_{i,0}$,
we find the "wavefunction" of the quasiparticle excitation, which is adiabatically connected to a boson insertion or removal at site $0$ in the limit $t=0$. This is easily seen from a Lehman decomposition of the Green's function.
%**
Note that it would be highly non-trivial to derive the result (\ref{result}) from a naive perturbation theory for $G^R_{i,0}$. This may be appreciated from Fig.~\ref{Fig:Interference} which illustrates the forward scattering and its many-particle nature on a selected path. Remarkably, the exponentially many virtual trajectories between initial and final state sum up to the single product in Eq.~(\ref{result}). 
 
At low temperatures, transport of bosons is expected to proceed via variable range hopping, which is very sensitive to the localization length $\xi$ of excitations. For non-interacting fermions it is defined as the (log-averaged) inverse spatial decay rate of single particle wavefunction amplitudes. For hard core bosons, $\xi$ is naturally generalized to be the typical inverse decay rate of $G^R_{i,0}$ with distance, 
\bea
\label{loclength}
1/\xi(\e_0) = -\lim_{\vec{r}_i\to \infty}  \frac{\overline{\ln |G^R_{i,0}(\omega)/G^R_{0,0}(\omega)|}_{\omega\to \e_0}}{|\vec{r}_i-\vec{r}_0|},
\eea
the overbar denoting disorder average. Setting $\omega\to \e_0$ selects the decay rate of the excitation centered at site $0$. %with energy $\e_0$. 
 
Note that setting $B=0$, Eq.~(\ref{result}) reproduces the well-known result for non-interacting fermions~\cite{Anderson,NSS}, %MedinaKardar} 
which can also be extended to repulsive interactions~\cite{SpivakShklovskiiReview}. In contrast,
hard core bosons differ crucially in the sign of the amplitude contributed by the paths.
%The result is conveniently rewritten as
%\bea
%\label{resultbos}
%\frac{G^{R,{\rm bos}}_{i,0}(\omega)}{G^{R,{\rm bos}}_{0,0}(\omega)} %&=& 
%%\sum_{{\cal P}=\{j_0=0,j_1,...,j_\ell=i \}} \prod_{p=1}^{\ell} 
%%\frac{V_{j_p,j_{p-1}} (-{\rm sign}(\e_{j_p})) }{E-\e_{j_p}}\nn\\
%&=& \sum_{{\cal P}=\{j_0=0,...,j_\ell=i \}} \prod_{p=1}^{\ell} \frac{t_{j_{p-1},j_p}}{|\e_{j_p}|-{\rm sgn}(\e_{j_p}) \omega}.
%\eea
The difference is easy to understand, cf.~Fig.~\ref{Fig:Interference}. In order to observe a particle at site $i$ after inserting a particle at $0$, all the $n_{\cal P} \equiv \sum_{k=1}^\ell  n_k \approx \sum_{k=1}^\ell  [1-{\rm sgn}(\e_k)]/2$ particles on the path ${\cal P}$ have to move to the next negative energy site closer to site $i$. Upon retrieving a particle at site $i$, a ring exchange of $n_{\cal P}$ particles has been  carried out in the ground state, which causes the sign difference $(-1)^{n_{\cal P}}$ between bosonic and fermionic amplitudes.
%This  may have been guessed on the basis of a Jordan-Wigner transform along a path ${\cal P}$.   

In the impurity band model of Eq.~(\ref{H}), this feature distinguishes clearly between bosons and fermions. However, we should mention that sums over positive paths can also occur in fermionic problems~\cite{ES_MR}. Such situations arise, when all sites between $0$ and $i$ have energies above the considered $\omega$. This occurs, e.g., in lightly doped semiconductor solutions, where impurity states tunnel through the bottom of the disordered conduction band; or in impurity bands with chemical potential very close to the band edges.%~\cite{fnbandedge}

{\em Effects of quantum statistics and magnetoresistance -} 
Eq.~(\ref{result}) shows that for low energy bosonic excitations, $\omega \to 0$, in the absence of a magnetic field, all paths interfere constructively, unlike in typical fermionic situations. This may be seen as a precursor of the establishment of global phase coherence in a superfluid phase. 
A simple consequence of this difference is that in the same disorder potential hard core bosons always have larger localization length than fermions of the same mass. 

The difference in path signs has also a crucial effect on magnetoresistance, as was also noted (for $\omega=0$) in Refs.~\cite{Zhao,Syzranov}.
% and will shed new light not only on strongly localized insulators, but also on the approach to superfluidity.
It manifests itself prominently in a strong opposite response to a magnetic field $H$ depending on the statistics of carriers. It is well known that hopping fermions experience an (exponentially strong) negative magnetoresistance due to the suppression of destructive interference~\cite{NSS,SivanImry, MedinaKardar}. 
%The latter arises because the magnetic field renders the path amplitudes complex, which makes it harder for the paths to interfere completely destructively. This tends to decrease the large resistances in the hopping network and enhances hopping transport. 
In contrast, the magnetoresistance of bosons is positive, since the phases in the hopping amplitudes reduce the constructive interference of paths that connect the low energy sites relevant for transport. 
%**
%The replica scaling arguments of Ref.~\cite{MedinaKardar}, which maps the forward scattering problem to directed polymers, should apply also to the bosonic case. 
As long as the relevant hopping distance $r=R_{\rm hop}$ is small, magnetoresistance is weak, since only a small fraction of a flux quantum threads the wavefunction on that scale. However, while fermions react with a non-analytic increase $\Delta \overline{\log{G}}\sim |H|$ (due to destruction of nearly perfect negative interference of competing paths~\cite{NSS}), bosons display a smaller, analytic response of opposite sign,  $\Delta \overline{\log{G}}\sim - p_R \Phi_R^2\sim H^2 R_{\rm hop}^3$. Here, $\Phi_R\sim H R^{1+\zeta}$ is the flux through a pair of typical tunneling paths of length $R$, while $p_R\sim R^{-\theta}$ is the probability for two paths to interfere significantly. $\zeta\geq 1/2$ is the wandering or roughness exponent of directed paths in a disordered environment, while the exponent $\theta=2\zeta-1$ follows from a standard scaling relation~\cite{HalpinHealy}. In the earlier literature on fermion scattering problems with positive amplitudes~\cite{ES_MR}, the  result   $\Delta \overline{\log{G}}\sim H^2 R_{\rm hop}^3$  was obtained for the special case of weak disorder where the paths behave like random walks with $\zeta=1/2$.  

For larger $R_{\rm hop}$ or stronger $H$~\cite{fncrossover}, the interference effects actually {\em shrink} the boson localization length, in analogy to the opposite effect in fermions, i.e., $\xi(H)-\xi(0)\propto (-1)^B H^\alpha$~\cite{Zhao,KardarBook}. This leads to {\em exponentially amplified, giant magnetoresistance} in the low temperature hopping regime~\cite{DPRM_MR}, where the hopping length is proportional to an inverse power of $T$. Hence, under a magnetic field, the typical hopping resistance $\sim\exp[R_{\rm hop}/\xi(B)]$ increases by a large factor, while the resistance of fermions typically decreases by a (significantly smaller) factor. 
%might be dropped
%However, we see no reason to expect activated transport, which was conjectured in Ref.~\cite{Syzranov}. 

The opposite interference in bosons and fermions is very likely to be a key element for understanding the giant magnetoresistance peak in disordered films with remnant pairing. As long as the magnetic field does not destroy the localized pairs, it mainly reduces their localization length. 
Upon destruction of the pairs, e.g., by the Zeeman effect, the predominant carriers are fermions, for which a negative magnetoresistance due to an increasing localization length is predicted~\cite{NSS,MedinaKardar}. Once the latter becomes large, the physics of loops (neglected in the forward scattering approximation) is likely to play a role in the negative magnetoresistance, as well. In this regime, effects of Coulomb interactions~\cite{Anirban}, and the necessity of electrons to tunnel around or  through remnant superconducting islands may enhance the negative magnetoresistance even further.~\cite{Meir} 

Note that the mechanism of positive magnetoresistance discussed above is based on transport via purely bosonic  carriers (pairs of electrons). This differs from other theoretical scenarii~\cite{Meir} where the bottleneck of resistance is due to the transfer of single electrons between remnant superconducting islands, and the positive magnetoresistance is ascribed to the shrinking of those islands. 

Purely bosonic transport is suggested by recent experiments on periodically patterned films of Bi or InO$_x$~\cite{Valles, LittleParks}. Indeed, the observed oscillations of  magnetoresistance start with an upturn, as expected for bosons, in contrast to the downturn characteristic for fermions. More importantly yet, the oscillations come with a flux periodicity corresponding to carriers with charge $2e$, suggesting that transport is carried by "pairs" of electrons. 
%Such a feature was indeed observed in Bi films~\cite{Valles}, and may be taken as a hint for bosonic carriers in those systems.
 
\begin{figure}[h]
\centerline{\includegraphics[width=0.48\textwidth]{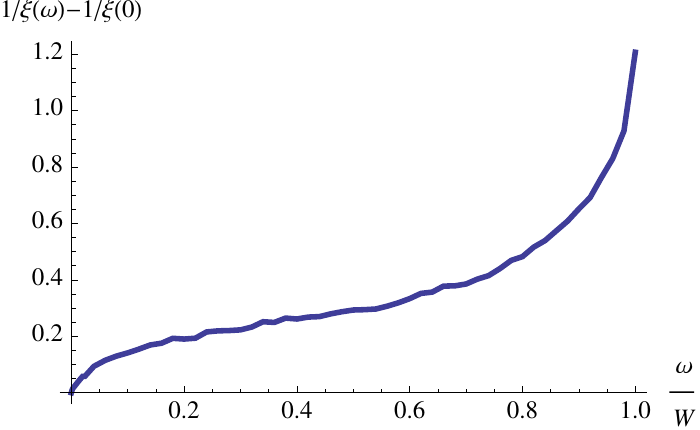}}
\caption{Disorder averaged spatial decay rates $\xi^{-1}$ of bosonic excitations along the diagonal of a square lattice. $\xi$ is computed from Eqs.~(\ref{result},\ref{loclength}) as a function of energy $\omega$ and measured in units of inverse lattice spacing. Note that the excitations of lowest energy are the most delocalized.}
\label{fig:xi}
\end{figure}

{\em Energy dependence of  $\xi$ - } An interesting consequence of Eq.~(\ref{result}) is the prediction that for bosons $\xi(\omega)$ has a non-trivial energy dependence around $\omega=0$. Indeed it reaches a {\em maximum} at $\omega=0$, as we confirmed numerically in Fig.~\ref{fig:xi}. The presence of other bosons thus enhances the delocalization tendency of an extra particle at low energy, in contrast to non-interacting fermions which are essentially insensitive to the position of the Fermi level. 
Note that at higher energies %$\omega\sim W$, 
bosonic excitations tend to behave like non-interacting particles, since paths through occupied sites become negligible. 
%These high energy excitations are however irrelevant for d.c. transport.

So far the above discussion of $\xi(\omega)$ was based on the forward scattering approximation, which yields the recursive relation (\ref{recursion}) between Green's functions at the {\em same} $\omega$, almost like in a non-interacting problem. This observation can be used to define an effective single particle Hamiltonian with complex hopping amplitudes $t_{ij} = t [{\rm sign}(\epsilon_i\epsilon_j)]^{1/2}$, which generates the same expression as in the interacting hard core boson problem for {\em all}, not only the shortest, non-intersecting paths. The study of such effective non-Hermitian Hamiltonians is an interesting subject for future studies.

%The decrease of $\xi(\omega)$ with $\omega$ is predicted to leading order in the hopping, which certainly applies to intermediate scales where forward scattering dominates the propagation. 
{\em Approach to superfluidity and delocalization - } The predicted decrease of $\xi(\omega)$ with increasing $\omega$ 
%It will be interesting to understand whether such a feature survives when subleading terms and loop corrections are added. It 
may appear counterintuitive at first sight, since at higher energies more phase space is available, which generally favors delocalization. However, one should interpret the phenomenon of a decreasing $\xi(\omega)$ as a precursor of incipient long range order, which will eventually establish at $\omega=0$, and favors propagation at low frequencies in local "precondensates". 
In the closely related random transverse field Ising model, cf.~Eq.~(\ref{Ising}) below, exact results for localization properties are available in 1d, due to the mapping to free fermions. Those exhibit indeed the same qualitative behavior of $\xi(\omega)$.~\cite{BouchaudLeDoussal} 
%The phenomenology obtained within the forward scattering approximation is closely analogous to that of phonons, whose localization length increases with decreasing energy, as well. 
%Similarly, the density of states of fermions is hardly affected by weak hopping, since self-energy corrections to the on-site energies (from loops renormalizing the forward scattering~\cite{Anderson}) have arbitrary signs. In contrast, the same kind of corrections for bosons systematically decrease the absolute value of effective site energies, $\e_i^{\rm ef} =\e_i+\Sigma_i(\omega\approx 0)$, with the self-energy $\Sigma_i = -{\rm sgn}(\e_i) \sum_{j\in \partial i }t_{ij}^2/|\e_j|+O(t^4)$
%as $\e_i \Sigma_i <0$ for low energy processes. This suggests an increased density of states at lowest energy, which may be observable in spectral functions or the specific heat. 
%Using as an estimate for the superfluid transition point in 3d the condition that the Green's function evaluated at $\omega$ do not decay exponentially (and thus the locator expansion diverges), yields $(W/t)_c\approx 36$. Including self-energy effects via exclusion of small denominators $<\alpha t^2/W$ (Anderson's upper limit criterion) we find $(W/t)_c\approx27$, as opposed to the value of $18.5$ known rather precisely for non-interacting fermions. We chose $\alpha=7$ instead of Anderson's suggestion of $4$, to reproduce the correct critical value for fermions.  
These results contradict the predictions of Refs.~\cite{Mueller09, IoffeMezard}, which argued that the presence of a sea of hard core bosons  impedes the propagation of an extra boson injected at low energy.~\cite{fnIM}
However, while this reasoning would be correct for a {\em distinguishable} extra particle, it neglects  exchange effects of identical bosons, which instead lead to {\em enhanced} propagation at low energies. 
%The competition with the opposite trend of enhanced delocalization due to larger phase space will be discussed further below.

The locator expansion is helpful to understand qualitatively another aspect of bosonic (de)localization: How do bosons escape localization in $d=2$, while repulsive fermions are believed to always localize in the absence of special symmetries?
Usually one argues that superfluids in 2d are stable to weak disorder, which proves their delocalization~\cite{LeeGunn}. The approach of this work complements this view from the insulating side. At low energies all scattering paths interfere constructively, which is a precursor of the establishment of a global phase in the superfluid. This is very different from fermions where the various scattering paths have nearly random signs, such that quantum interferences essentially average out, {\em except} for paths returning back to the origin.  For the latter, "time reversed paths" (i.e., sequences of scattering states encountered in perturbation theory, and their reverse) are guaranteed to have the same scattering amplitude in the absence of magnetic fields. Their positive interference thus systematically enhances the return to the origin and therefore localization. In contrast, boson propagation at low energy {\em always} involves positive interference of alternative paths, such that the return to the origin is not particularly enhanced as compared to other propagation channels.   
%, which guarantee constructive interference, enhance the return probability and eventually localize the fermions, 
%appear unimportant for bosons, since most paths between {\em any} two points interfere constructively, not only those with coinciding initial and final point.

%\bibitem{fnIM} The calculation of Ref.~\cite{IoffeMezard} arrived at such a result by an unjustified restriction of the perturbation series of $G^R_{i0}(\omega)$  to virtual one-site excitations, instead of summing the exponentially many terms corresponding to all orders in which the many particle rearrangments in Fig.~\ref{} can be made. 

Let us now attempt to obtain insight on the approach to superfluidity. We may use the locator expansion technique to revisit the problem of hard core bosons on Cayley trees of large connectivity $K\gg 1$, as considered in Ref.~\cite{IoffeMezard}. Such high connectivity lattices are indeed interesting since they enable one to use the forward scattering approximation even parametrically close to the superfluid transition. They thus yield insight into how bosonic excitations approach delocalization, and how this differs from the exactly solvable case of free fermions~\cite{AbouChacra}.   

In finite dimensions, superfluidity sets in when $\xi(\omega=0)$ diverges. 
%At that point the bosons to condense into a delocalized collective state developping at the chemical potential and to form a superfluid.~\cite{fn}  
On the Cayley tree, the criterion generalizes to $\xi^{-1}_{\omega=0}=\lim_{R\to \infty}R^{-1} \overline{\ln[\sum_{i,{\rm dist}(i,0)=R} {G^R_{i0}(\omega=0)}]}=0$. This can be evaluated by a mapping to a directed polymer~\cite{IoffeMezard, DerridaSpohn}, which is exact within the forward scattering approximation.
Due to the absence of loops, any two sites are connected by a unique shortest path. Hence, interference phenomena are subleading in the hopping. To leading order quantum statistics is therefore irrelevant, and one finds localization properties like for free fermions (as characterized by $|G^R_{i0}(\omega)|^2$ at large distances) and a superfluid transition at the same value at which non-interacting fermions delocalize, $(t/W)_c = O(1/K\ln(K))$~\cite{AbouChacra}. 
%footnote?
%{This is ensured by the fact that the sums $\sum_{i,{\rm dist}(i,0)=R\to \infty} |G^R_{i0}(\omega=0)|^\alpha$ ($\alpha=1,2$) are both dominated by only a few rare boundary sites $i$, whose number remains small as $R\to \infty$~\cite{IoffeMezard}. Thus they diverge at the same value $t/W$.}
 However, a study of subleading corrections shows that bosons actually delocalize already at a weaker hopping strength than fermions~\cite{Bapst}.  Like for the critical wavefunctions of the fermionic problem, one finds that the emerging Bose condensate is extremely sparse, as pointed out in Ref.~\cite{IoffeMezard}.
    
%The appearance of a condensate at $T=0$ and the delocalization of excitations at lowest energy are dominated by the same rare paths. At the transition, the bosons are thus expected to condense into the delocalized state occurring at $\omega=0$.  Neglecting self-energy and other corrections of higher order in $(t/W)$, the wavefunction of the condensate corresponds to a critical wavefunction at the single particle Anderson transition, and thus inherits the strong fractality of the latter. It seems natural to assume that the fractality of the critical condensate persists also upon including higher order corrections. 

In the insulator, the leading order locator expansion shows that the typical propagator at finite energies, $G^R_{i0}(\omega>0)$, always decays faster than $G^R_{i0}(0)$, if a uniform distribution of random energies ($\rho(\epsilon\approx \mu)={\rm const.}$) and chemical potential $\mu=0$ (half filling) is assumed. However, the range of $\omega$, for which this leading order result is controlled, gradually decreases to zero upon approaching the phase transition. A complete description of criticality at finite energies would require the resummation of very high orders of perturbation theory. Nevertheless, this result
is suggestive of the possibility that the superfluid emerges out of the insulator by a delocalization phenomenon at $\omega=0$, while slightly higher (intensive) excitations are still localized -- a scenario which we indeed find below for Ising models.
This contrasts with the scenario of a mobility gap in the insulator, that closes at the transition, as proposed in~\cite{Mueller09,IoffeMezard}, and similar early ideas by Hertz et al.~\cite{HertzAnderson}. All of those neglected the above discussed exchange effects of identical particles at low energies.  
%Since effects beyond the forward scattering approximation reinforce this tendency~\cite{Bapst}, and since the interference phenomena in finite dimensions work in the same direction (cf.~Fig.~\ref{fig:xi}), we conjecture that the delocalization initiated at $\omega=0$ holds quite generally if the distribution of onsite disorder is constant around $\epsilon=\mu$. 
%Indeed, this agrees with what one would expect from a strong randomness approach (cf. e.g. the review~\cite{Vojta}) where higher energy excitations are more localized. 
%On the other hand, our result 

The present calculations do not provide any evidence for such a mobility edge at higher (intensive) energies in the presence of uniform disorder.
However, an intensive mobility edge, and even a closing mobility gap, does arise rather trivially if the density of on-site energies increases sufficiently strongly across the chemical potential. In such cases the forward scattering approximation on the Cayley tree indeed predicts an intensive mobility edge that closes at criticality, very much like at a standard Anderson transition of fermions. %However, the increase of the density of states needs to be strong enough to compete with both the above mentioned mechanisms which favor the opposite behavior, especially in finite dimension. 

%Let us finally address the question of localization at higher energies. The lowest energy excitations above the ground state should be well captured by the first few terms in the locator expansion, including the most important subleading terms~\cite{Bapst}, and thus allow us to conclude about the localization behavior at very low energies. However,  this is not necessarily true for higher energies where multi-particle-like excitations may propagate by continuously scattering from each other. Such effects are, however, hard to detect at large connectivity $K$, where collisions of single-particle-like excitations are very inefficient in enhancing delocalization~\cite{IoffeMezard}. It might even be that in this limit the insulating phase of the model (\ref{H}) on the Bethe lattice is fully localized, even at extensive energies. However, such an abrupt jump from a superfluid to a complete "superinsulator" is most likely an artifact of the large $K$ limit, whereas at finite connectivity an intermediate regime with delocalized high energy excitations is expected. Those may arise either at extensive energies only, as conjectured in Ref.~\cite{Aleiner09} for low dimensions $d\leq 2$, or already at intensive (finite) energies, which is certainly possible in higher dimensions.  Analytically, however, the study of this question would require the analysis of very high orders of subleading terms in the $1/K$ expansion. 

{\em Locator expansion for Ising spins -} Some of the qualitative features found for hard core bosons  also hold for the closely related random transverse field Ising model, 
%(with $s=1/2$ spins), 
\bea
\label{Ising}
H_{\rm Ising} = \sum_i \e_i s^z_i 
- 4 \sum_{\langle i,j\rangle }  t_{ij} s^x_i s^x_j,
\eea
even though its critical behavior turns out to be rather different.~\cite{Xiaoquan}
This model differs from the disordered XY model of Eq.~(\ref{H}) only by the replacement $2(s^x_i s^x_j+ s^y_i s^y_j)\to 4 s^x_i s^x_j$ (after applying to Eq.~(\ref{H}) the standard mapping between hard core bosons and $s=1/2$ spins~\cite{footnote:spinhalf}). 
In this case, analogous steps as in Eqs.~(\ref{Eqmot}-\ref{result}), applied to correlators $\langle [s^x_i(t), s^x_0(0)]\rangle$, yield a sum over shortest paths, with amplitudes given by products of factors $2|\e_i|/(\e_i^2-\omega^2)$, that replace the XY locator ${\rm sign}(\e_i)/(\e_i-\omega)$ in Eq.~(\ref{result}). This differs from the factor $2/(|\e_i|-\omega)$ postulated in Refs.~\cite{IoffeMezard},  which incorrectly predicted an intensive mobility edge that closes at the transition point. Taking into account the effects of higher order terms in the expansion in exchange coupling and the special role of rare events in Ising models, which are well known from 1d chains and strong randomness approaches~~\cite{Vojta, Kovacs,BouchaudLeDoussal,Xiaoquan}, we find instead that the ordering transition is initiated by a delocalization at $\omega=0$, while slightly higher-lying intensive excitations are still localized in the paramagnet. On the other hand, we cannot reliably analyze high but finite energies, as this requires full control over very high orders of perturbation theory.  

Ref.~\cite{NatCom} attempted to address the question of a mobility edge in the disordered phase. Numerically studying certain Ising models on small random graphs with $K=2$, the authors claimed to find a mobility edge at intensive energies~\cite{fnRandomgraphs}. However, it remained unclear whether this mobility edge was actually found on the paramagnetic side of transition, and if so, whether the mobility gap remains finite at the phase transition (as predicted here).  Similar studies in finite dimensional Ising and XY models would therefore be very desirable to determine whether also there delocalization is  initiated at $\omega=0$ in Ising models, and whether the same holds for XY models. 

%However, unfortunately, random graphs are plagued by the problem that the delocalization of excitations does not coincide with a change in level statistics (targeted in Ref.~\cite{NatCom}) as was recently shown~\cite{BiroliTarzia},  Furthermore, random graphs suffer from logarithmic finite size effects, which are partly responsible for the large error bars in Ref.~\cite{NatCom}. 

%In more general models the Bose glass adjacent to the superfluid will not be a fully localized insulator at all energies of order $O(1)$.  Instead, it may possess a finite, albeit non-critical (non-closing) mobility gap, which can give rise to activated or superactivated transport~\cite{Mueller09}. Such a situation arises rather trivially if the bare density of states of the disorder potential, $\rho(\e)$ increases with $\e>\mu$, since excitations of higher energy hybridize more easily with neighboring sites than lower excitations. A less trivial, but still solvable model, where an inhomogeneous density of states is self-generated will be presented elsewhere~\cite{Xiaoquan}.  

{\em Conclusion -} We have shown that strongly disordered bosons respond oppositely to a magnetic field than fermions in impurity bands, which makes magnetoresistance a measurement of choice to detect the statistics of the charge carriers in an insulator.  
We hope that the non-trivial dependence of the localization length $\xi(\omega,H)$ on energy, magnetic field and statistics will be studied in superconducting films or in cold bosonic atoms~\cite{Roati}, where artificial "magnetic" gauge fields can be generated by various techniques~\cite{Dalibard}. The positivity of bosonic tunneling amplitudes at zero energy furnishes an intuitive understanding of why bosons escape localization in 2d.  
Applying the locator expansion to Ising systems as well, we predict that long range ordered phases may emerge from quantum  disordered phases by delocalizing and condensing at $\omega=0$, without the closure of a pre-existing intensive mobility edge.
%We hope to push the locator expansion to approach the superfluid-insulator transition in finite dimensions.

I would like to thank V. Bapst, V. Kravtsov, S.V. Syzranov and X. Yu for useful discussions.

%\vspace{-.5 cm}


\begin{thebibliography}{99}
\bibitem{fncrossover} The crossover happens when $H r^{1+\zeta}=O(1)$, contrary to assumptions in the earlier literature~\cite{KardarBook}, which predicting it at  $H r^{3/2}=O(1)$ viewing paths as random walks with $\zeta=1/2$. This is, however, not appropriate in strong disorder.~\cite{DPRM_MR}
\bibitem{fnIM} Ref.~\cite{IoffeMezard} arrived at such a result by restricting the perturbation series of $G^R_{i0}(\omega)$  to virtual one-site excitations, instead of summing the exponentially many ways in which a minimal number of sequential moves lead to the many-particle rearrangment in Fig.~\ref{Fig:Interference}.
\end{thebibliography}
\end{document}